\documentclass[journal]{IEEEtran}

\ifCLASSINFOpdf
\else
   \usepackage[dvips]{graphicx}
\fi
\usepackage{url}

\usepackage{graphicx}

\usepackage{cite}
\usepackage{amsmath,amssymb,amsfonts}
\usepackage{algorithmic}
\usepackage{textcomp}
\usepackage{xcolor}
\def\BibTeX{{\rm B\kern-.05em{\sc i\kern-.025em b}\kern-.08em
		T\kern-.1667em\lower.7ex\hbox{E}\kern-.125emX}}

\usepackage{times,bm}
\usepackage[printonlyused, nolist]{acronym}
\usepackage{pgfplots}
\usepackage{booktabs}
\usepackage{pifont}
\usepackage{algorithm}

\usepackage{tikz}
\usetikzlibrary{plotmarks}
\tikzset{every picture/.style={line width=1.5pt}}

\pgfplotsset{compat=1.11}
\newlength\fwidth

\newcommand{\W}{\mathbf{W}}
\newcommand{\w}{\mathbf{w}}

\newcommand{\y}{\mathbf{y}}
\newcommand{\x}{\tilde{\mathbf{x}}}
\newcommand{\xWhite}{\mathbf{x}}

\newcommand{\V}{\mathbf{V}}

\newcommand{\WhiteningMat}{\mathbf{Q}}

\newcommand{\UpperBound}{U}

\newcommand{\FreqIdx}{f}
\newcommand{\FreqIdxMax}{F}
\newcommand{\BlockIdx}{n}
\newcommand{\BlockIdxMax}{N}
\newcommand{\ChannelIdx}{k}
\newcommand{\ChannelIdxMax}{K}

\newcommand{\yFreqVec}{\underline{\y}}

\newcommand{\SetW}{\mathcal{W}}

\newcommand{\transp}{^\text{T}}
\newcommand{\herm}{^\text{H}}

\newcommand{\Expect}[1]{\hat{\mathbb{E}}\left\lbrace #1 \right\rbrace}
\newcommand{\ExpectIdeal}[1]{\mathbb{E}\left\lbrace #1 \right\rbrace}

\newcommand{\Walter}[1]{#1}
\newcommand{\WalterTwo}[1]{#1}
\newcommand{\WalterThree}[1]{#1}
\newcommand{\Waspaa}[1]{#1}
\newcommand{\WaspaaTwo}[1]{#1}

\begin{document}

\title{Faster IVA: Update Rules for\\ Independent Vector Analysis based on\\ Negentropy and the Majorize-Minimize Principle}

\author{Andreas Brendel, \IEEEmembership{Student Member, IEEE}, Walter Kellermann, \IEEEmembership{Fellow, IEEE}
\thanks{This work was funded by the Deutsche Forschungsgemeinschaft (DFG, German Research Foundation) -- 282835863 -- within the Research Unit FOR2457 ``Acoustic Sensor Networks''.}
\thanks{Andreas Brendel and Walter Kellermann are with the chair of Multimedia Communications and Signal Processing, Friedrich-Alexander-Universit\"at Erlangen-N\"urnberg,
	Cauerstr. 7, D-91058 Erlangen, Germany,
	e-mail: \texttt{\{Andreas.Brendel, Walter.Kellermann\}@FAU.de}}}

\markboth{ }
{Brendel and Kellermann: Faster IVA: Update Rules for Independent Vector Analysis based on Negentropy and the Majorize-Minimize Principle}
\maketitle

\begin{acronym}
	\acro{STFT}{Short-Time Fourier Transform}
	\acro{PSD}{Power Spectral Density}
	\acro{PDF}{Probability Density Function}
	\acro{RIR}{Room Impulse Response}
	\acro{FIR}{Finite Impulse Response}
	\acro{FFT}{Fast Fourier Transform}
	\acro{DFT}{Discrete Fourier Transform}
	\acro{ICA}{Independent Component Analysis}
	\acro{IVA}{Independent Vector Analysis}
	\acro{TRINICON}{TRIple-N Independent component analysis for CONvolutive mixtures}
	\acro{FD-ICA}{Frequency Domain ICA}
	\acro{BSS}{Blind Source Separation}
	\acro{NMF}{Nonnegative Matrix Factorization}
	\acro{MM}{Majorize-Minimize}
	\acro{MAP}{Maximum A Posteriori}
	\acro{RTF}{Relative Transfer Function}
	\acro{AuxIVA}{Auxiliary Function IVA}
	\acro{FD-ICA}{Frequency-Domain Independent Component Analysis}
	\acro{DOA}{Direction of Arrival}
	\acro{SNR}{Signal-to-Noise Ratio}
	\acro{SIR}{Signal-to-Interference Ratio}
	\acro{SDR}{Signal-to-Distortion Ratio}
	\acro{SAR}{Signal-to-Artefact Ratio}
	\acro{GC}{Geometric Constraint}
	\acro{DRR}{Direct-to-Reverberant energy Ratio}
	\acro{ILRMA}{Independent Low Rank Matrix Analysis}
	\acro{IVE}{Independent Vector Extraction}
	\acro{GC-IVA}{Geometric Constraint IVA}
	\acro{SOI}{Sources Of Interest}
	\acro{BG}{Background}
	\acro{MNMF}{Multichannel NMF}
	\acro{IP}{Iterative Projection}
	\acro{EVD}{Eigenvalue Decomposition}
	\acro{GEVD}{Generalized Eigenvalue Decomposition}
	\acro{SQUAREM}{Squared Iterative Methods}
	\acro{EM}{Expectation Maximization}
	\acro{PCA}{Principal Component Analysis}
	\acro{RV}{Random Vector}
\end{acronym}

\begin{abstract}
\Waspaa{Algorithms for} Blind Source Separation (BSS) of acoustic signals \Waspaa{require} efficient and fast converging optimization strategies to adapt to \Waspaa{nonstationary signal statistics and time-varying acoustic scenarios}. In this paper, we \Waspaa{derive} fast converging update rules \Waspaa{from a negentropy perspective, which are} based on the Majorize-Minimize (MM) principle and eigenvalue decomposition. The presented update rules are shown to \Waspaa{outperform competing state-of-the-art methods} in terms of convergence speed at a comparable runtime \Waspaa{due to the restriction to unitary demixing matrices. This is} demonstrated by experiments with recorded real-world data.
\end{abstract}

\begin{IEEEkeywords}
Independent Vector Analysis, fast convergence, MM Algorithm, FastIVA
\end{IEEEkeywords}

\IEEEpeerreviewmaketitle

\section{Introduction}
\label{sec:intro}
\ac{BSS} aims at separating sources from an observed mixture by using only very weak assumptions about the underlying scenario. Hence, such methods are applicable in a variety of situations \Walter{\cite{hyvarinen_independent_2001,comon_handbook_2010,makino_blind_2007,vincent_audio_2018}}. One important aspect in the design of \ac{BSS} algorithms is the development of fast converging and at the same time computationally simple optimization strategies. For \ac{ICA} the \Walter{FastICA} update rules based on a fixed-point iteration scheme represent the gold standard in this research field \cite{hyvarinen_fast_1999}. This update scheme is derived by maximizing the so-called negentropy, i.e., by maximizing the nongaussianity of each separated source. Several variants of these updates including the extension to complex-valued data have been proposed \cite{bingham_fast_2000}.

In this contribution, we consider mixtures of acoustic sources \cite{makino_blind_2007,vincent_audio_2018}. The most important difference of \ac{BSS} methods for acoustic mixtures \WalterTwo{relative} \Walter{to instantaneous problems \cite{hyvarinen_independent_2001}} is the mixture model: Observed acoustic signals undergo propagation delay and multipath propagation. Hence, a convolutive mixture model is needed. A well-established \Walter{concept} is to transform the problem into the \ac{STFT} domain and solve instantaneous \ac{BSS} problems in each frequency bin \cite{smaragdis_blind_1998}. However, this causes the well-known inner permutation problem which has to be resolved \Walter{by additional heuristic measures} to obtain decent results \cite{sawada_robust_2004}. As an alternative which aims at avoiding the inner permutation problem, \ac{IVA} has been proposed \cite{kim_blind_2007}. A fast fixed-point algorithm called \Walter{FastIVA} has been developed following the ideas of \Walter{FastICA} \cite{hyvarinen_fast_1999} for the optimization of \ac{IVA} \cite{lee_fast_2007}. Fast and stable update rules have been developed based on the \ac{MM} principle and the iterative projection technique \cite{ono_stable_2011} and methods for accelerating its convergence have been investigated \cite{brendel_accelerating_2021}. Even faster update rules for the specific case of two sources and two microphones based on a \ac{GEVD} have been presented in \cite{ono_fast_2012}.

For source extraction \WaspaaTwo{\cite{brendel_journal,koldovsky_dynamic_2021}}, i.e., the separation of a desired source from a set of \Walter{multiple} interfering sources, update rules based on an \ac{EVD} of a weighted microphone covariance matrix have been proposed \cite{scheibler_2020} and spatial prior knowledge about the source of interest has been introduced in these update rules in \cite{brendel_eusipco}. \WaspaaTwo{Recently, priors on the source signal spectra for \ac{IVA} were proposed based on deep neural networks  \cite{makishima_independent_2019,nugraha_flow-based_2020}.}

In this contribution, we propose a new update scheme for \ac{IVA} described in terms of the negentropy of the demixed signals and based on the \ac{MM} principle. The optimization of the upper bound of the \ac{MM} algorithm is posed as an eigenvalue problem, which allows for fast convergence of the algorithm. In comparison to \cite{ono_fast_2012}, our \ac{EVD}-based update scheme allows for the separation of an arbitrary number of sources instead of only two. In \cite{scheibler_2020}, a structurally similar update scheme has been derived for the extraction of a single source from a different perspective. Here, we derive update rules which are also capable of separating an arbitrary number of sources. \Waspaa{Update} rules for \Waspaa{extracting a single source} are included in the proposed method as a special case. \Waspaa{We note that} \Walter{FastIVA} \cite{lee_fast_2007} uses \Walter{the same} cost function but uses a fixed-point algorithm for its optimization. The superiority of our proposed method \Walter{over FastIVA and AuxIVA in terms of convergence speed and separation performance after convergence} is demonstrated by experiments \Waspaa{using} real-world data created from measured \acp{RIR}.
%
\section{Cost Function}
\label{sec:cost_function}
We consider a determined scenario, in which $\ChannelIdxMax$ source signals are captured by $\ChannelIdxMax$ microphones with microphone signals described in the \ac{STFT} domain as
\begin{equation}
\x_{\FreqIdx,\BlockIdx} := \left[\tilde{x}_{1,\FreqIdx,\BlockIdx},\dots,\tilde{x}_{\ChannelIdxMax,\FreqIdx,\BlockIdx}\right]\transp \in \mathbb{C}^{\ChannelIdxMax},
\end{equation}
where $\FreqIdx \in \{1,\dots,\FreqIdxMax\}$ \Waspaa{indexes} the frequency and \mbox{$\BlockIdx \in \{1,\dots,\BlockIdxMax\}$} the \Waspaa{time} frame. The aim of the developed algorithm is to estimate the demixed signals
\begin{equation}
\y_{\FreqIdx,\BlockIdx} := \left[y_{1,\FreqIdx,\BlockIdx},\dots,y_{\ChannelIdxMax,\FreqIdx,\BlockIdx}\right]\transp  \in \mathbb{C}^{\ChannelIdxMax}
\end{equation}
from the microphone signals $\x_{\FreqIdx,\BlockIdx}$. For notational convenience, we introduce the broadband vector of the demixed signal of channel $\ChannelIdx$ and time frame $\BlockIdx$
\begin{equation}
\yFreqVec_{\ChannelIdx,\BlockIdx} := \left[y_{\ChannelIdx,1,\BlockIdx},\dots,y_{\ChannelIdx,\FreqIdxMax,\BlockIdx}\right]\transp  \in \mathbb{C}^{\FreqIdxMax},
\end{equation}
which is modeled to follow a multivariate supergaussian \ac{PDF} $p(\yFreqVec_{\ChannelIdx,\BlockIdx})$, where all frequency bins are modeled to be uncorrelated but statistically dependent. Examples for such \acp{PDF}, which are typically used for \ac{IVA} include the multivariate Laplacian \ac{PDF} or the generalized Gaussian \ac{PDF} \cite{ono_auxiliary-function-based_2012}. 

\Waspaa{In the following, signal vectors without frame index $\BlockIdx$ denote \acp{RV} and signal vectors with frame index their realizations.} As the \ac{PDF} of a mixture of multiple \WaspaaTwo{independent non-Gaussian source signals} tends toward a Gaussian, maximizing the \emph{negentropy} \cite{hyvarinen_independent_2001} of the \Waspaa{\ac{RV}} of the demixed signals $\yFreqVec := [\yFreqVec_{1}\transp,\dots,\yFreqVec_{\ChannelIdxMax}\transp]\transp$ is an intuitive and widely used \ac{BSS} cost function. The negentropy, i.e., Kullback-Leibler divergence between the \acp{PDF} of the \Waspaa{\acp{RV}} $\yFreqVec$ and $\underline{\mathbf{z}}$, where the latter is normally distributed with same mean vector and covariance matrix as $\yFreqVec$, is defined by
\begin{align}
N(\yFreqVec) := \mathcal{KL}\left\lbrace p(\yFreqVec)\Vert p(\underline{\mathbf{z}}) \right\rbrace =  \mathcal{H}\left(\underline{\mathbf{z}}\right) - \mathcal{H}\left(\yFreqVec\right).
\label{eq:negentropy}
\end{align}
Note that the differential entropy of a Gaussian \Waspaa{\ac{RV}} $\mathcal{H}\left(\underline{\mathbf{z}}\right)$ represents a constant and is irrelevant for optimization. To ensure non-trivial solutions if multiple sources should be separated and to fix the scaling of the demixed signals, a common constraint is to impose whiteness of the demixed signals $\ExpectIdeal{\y_{\FreqIdx,\BlockIdx}\y_{\FreqIdx,\BlockIdx}\herm} = \mathbf{I}_\ChannelIdxMax$. Under this restriction, we obtain (cf. \cite{cover_elements_2006})
\begin{equation}
N(\yFreqVec) \geq \sum_{\ChannelIdx=1}^{\ChannelIdxMax} N(\yFreqVec_\ChannelIdx) = \sum_{\ChannelIdx=1}^{\ChannelIdxMax} \underbrace{\mathcal{H}\left(\underline{\mathbf{z}}_\ChannelIdx\right)}_\text{const.} - \mathcal{H}\left(\yFreqVec_{\ChannelIdx}\right),
\label{eq:channelWise_negentropy}
\end{equation} 
where $\underline{\mathbf{z}}_\ChannelIdx$ is \Waspaa{defined} analogously \Waspaa{to} $\underline{\mathbf{z}}$. In the following, we will consider the maximization of the sum of the channel-wise negentropies $N(\yFreqVec_\ChannelIdx)$ as a surrogate for the maximization of $N(\yFreqVec)$. The requirement $\ExpectIdeal{\y_{\FreqIdx}\y_{\FreqIdx}\herm} = \mathbf{I}_\ChannelIdxMax$ can be accomplished by whitening the observed signals
\begin{equation}
\xWhite_{\FreqIdx,\BlockIdx} := \WhiteningMat_\FreqIdx \x_{\FreqIdx,\BlockIdx} \quad \WalterTwo{\text{with}} \quad \Walter{\WhiteningMat_\FreqIdx := \left(\ExpectIdeal{\x_{\FreqIdx}\x_{\FreqIdx}\herm}\right)^{\text{-}\frac{1}{2}}}
\label{eq:whitening_trafo}
\end{equation}
and estimating the demixed signals
\begin{equation}
\y_{\FreqIdx,\BlockIdx} = \W_\FreqIdx \xWhite_{\FreqIdx,\BlockIdx},
\end{equation}
with a unitary demixing matrix (cf. \cite{hyvarinen_independent_2001})
\begin{equation}
\WalterThree{\W_\FreqIdx := \begin{bmatrix}
	\w_{1,\FreqIdx},\dots,\w_{\ChannelIdxMax,\FreqIdx}
	\end{bmatrix}\herm \in \mathbb{C}^{\ChannelIdxMax\times\ChannelIdxMax}}.
\label{eq:demixing_equation}
\end{equation}
Here, $\w_{\ChannelIdx,\FreqIdx}$ denotes the demixing filter which extracts the $\ChannelIdx$th source signal \Waspaa{sample $y_{\ChannelIdx,\FreqIdx,\BlockIdx}$ at frequency $\FreqIdx$ and time frame $\BlockIdx$}.
By using the definition of the differential entropy $\mathcal{H}(\cdot)$ and the source model $G$\WaspaaTwo{,}
\begin{equation}
\WaspaaTwo{\mathcal{H}(\yFreqVec_{\ChannelIdx}) := \mathbb{E}\lbrace G(\yFreqVec_{\ChannelIdx})\rbrace\quad\text{with}\quad G(\yFreqVec_{\ChannelIdx}) := - \log \WalterTwo{p(\yFreqVec_{\ChannelIdx})}}\WaspaaTwo{,}
\end{equation}
we obtain the following optimization problem \Waspaa{by assuming i.i.d. signal frames (cf. \cite{hyvarinen_independent_2001,lee_fast_2007})}
\begin{align}
\underset{\W_{\FreqIdx} \forall \FreqIdx}{\text{minimize}} &\quad \sum_{\ChannelIdx=1}^{\ChannelIdxMax} \Expect{G\left(\yFreqVec_{\ChannelIdx,\BlockIdx}\right)}\label{eq:IVA_cost_function}\\
\text{subject to}&\quad \W_\FreqIdx\W_\FreqIdx\herm = \mathbf{I}_\ChannelIdxMax \ \forall \FreqIdx. \label{eq:othogonality_constraint}
\end{align}
Here, \eqref{eq:IVA_cost_function} \Waspaa{reflects} the maximization of channel-wise negentropies \eqref{eq:channelWise_negentropy} and \eqref{eq:othogonality_constraint} realizes the unitarity constraint on the demixing matrices $\W_\FreqIdx$. In \eqref{eq:IVA_cost_function}, we introduced the approximation of the expectation operator by arithmetic averaging over all available time frames \mbox{$\Expect{\cdot} := \frac{1}{\BlockIdxMax}\sum_{\BlockIdx = 1}^{\BlockIdxMax}(\cdot)$}. 
The optimization problem of \eqref{eq:IVA_cost_function} and \eqref{eq:othogonality_constraint} is closely related to the \ac{IVA} cost function \cite{kim_blind_2007}
\begin{equation}
J_\text{IVA}(\SetW) := \sum_{\ChannelIdx=1}^{\ChannelIdxMax} \Expect{G\left(\yFreqVec_{\ChannelIdx,\BlockIdx}\right)} - 2\sum_{\FreqIdx=1}^{\FreqIdxMax}\log \vert  \det\W_\FreqIdx \vert,
\label{eq:costFun_rephrased}
\end{equation}
where $\SetW$ denotes the set of demixing vectors $\w_{\ChannelIdx,\FreqIdx}$ of all frequency bins $\FreqIdx$ and channels $\ChannelIdx$. The first term of the \ac{IVA} cost function \eqref{eq:costFun_rephrased} corresponds to \eqref{eq:IVA_cost_function}. The second term of \eqref{eq:costFun_rephrased} is a regularizer on the demixing matrices $\W_\FreqIdx$ ensuring \Waspaa{linearly} independent demixing filter vectors $\w_{\ChannelIdx,\FreqIdx}$. For unitary $\W_\FreqIdx$ this term is constant and, hence, is irrelevant for optimization. In the optimization problem of \eqref{eq:IVA_cost_function} and \eqref{eq:othogonality_constraint}, the role of the regularizer is taken by the (stronger) constraint of unitarity of~$\W_\FreqIdx$. However, the assumption of unitary demixing matrices is a significant restriction w.r.t. the \ac{IVA} cost function \Waspaa{as will become obvious in the experimental evaluations}.
%
\section{Update Rules}
\label{sec:updates}
In the literature, \Waspaa{predominantly} fixed-point algorithms (FastICA, FastIVA) have been used for the optimization of \Waspaa{negentropy-based \ac{BSS} cost functions} \cite{hyvarinen_independent_2001,bingham_fast_2000,hyvarinen_fast_1999,lee_fast_2007}. \Waspaa{Motivated by the success of \ac{MM}-based approaches \cite{hunter_tutorial_2004} for \ac{BSS} based on the minimum mutual information principle \cite{ono_stable_2011,ono_auxiliary-function-based_2012,kitamura_determined_2016}, we exploit the \ac{MM} principle for the optimization of the negentropy-based \ac{IVA} cost function \eqref{eq:IVA_cost_function}, \eqref{eq:othogonality_constraint} in this contribution.} 

In the following\Walter{,} $l \in \{1,\dots,L\}$ denotes the iteration index and $\SetW^{(l)}$ is the set of the $l$th iterates of all demixing vectors. 
The main idea of the \ac{MM} principle \cite{hunter_tutorial_2004} is to define an upper bound $\UpperBound$, which fulfills the properties of majorization \WalterThree{and tangency, i.e., equality iff $\SetW = \SetW^{(l)}$,}
\begin{equation}
\WalterThree{J(\SetW) \leq \UpperBound\left(\SetW\vert\SetW^{(l)}\right) \ \text{and} \ J\left(\SetW^{(l)}\right) = \UpperBound\left(\SetW^{(l)}\vert\SetW^{(l)}\right),}\notag
\end{equation}
w.r.t. the cost function $J$. The upper bound $\UpperBound$ should be designed such that its optimization is easier than the \Walter{iterative} optimization of the cost function \Walter{itself,} or\Waspaa{,} ideally\Waspaa{,} solvable in closed form. As minimization of the upper bound
\begin{equation}
\SetW^{(l+1)} =\underset{\SetW}{\text{argmin}} \ \UpperBound\left(\SetW\vert\SetW^{(l)}\right)
\label{eq:upperbound_minimization}
\end{equation}
\Walter{enforces monotonically decreasing values of $\UpperBound$}, the following 'downhill property' of \ac{MM} algorithms is obtained 
\begin{align}
J\left(\SetW^{(l+1)}\right) &\leq \UpperBound\left(\SetW^{(l+1)}\vert\SetW^{(l)}\right)\\
&\leq \UpperBound\left(\SetW^{(l)}\vert\SetW^{(l)}\right)=J\left(\SetW^{(l)}\right)\notag.
\end{align}
For the construction of the upper bound, we use the inequality \cite{ono_stable_2011} for supergaussian source models $\tilde{G}(r_{\ChannelIdx,\BlockIdx}) = G(\yFreqVec_{\ChannelIdx,\BlockIdx})$ dependent on the norm of the $\ChannelIdx$th demixed signal $r_{\ChannelIdx,\BlockIdx} := \Vert \yFreqVec_{\ChannelIdx,\BlockIdx}^{(l)} \Vert_2$
\begin{equation}
\Expect{\tilde{G}\left(r_{\ChannelIdx,\BlockIdx}\right)} \leq \frac{1}{2}\sum_{\FreqIdx=1}^{\FreqIdxMax}\w_{\ChannelIdx,\FreqIdx}\herm \V_{\ChannelIdx,\FreqIdx}\w_{\ChannelIdx,\FreqIdx} +\text{const.}
\label{eq:ono_inequality}
\end{equation}
Here, we introduced the weighted covariance matrix of microphone observations
\begin{equation}
\V_{\ChannelIdx,\FreqIdx} := \Expect{\frac{\tilde{G}'\left( r_{\ChannelIdx,\BlockIdx}\right)}{r_{\ChannelIdx,\BlockIdx}}\xWhite_{\FreqIdx,\BlockIdx}\xWhite_{\FreqIdx,\BlockIdx}\herm}.
\label{eq:weighted_covMat}
\end{equation}
Combining \eqref{eq:IVA_cost_function} and \eqref{eq:ono_inequality} and neglecting constant terms yields \Waspaa{a} surrogate for the optimization problem \eqref{eq:IVA_cost_function}, \eqref{eq:othogonality_constraint} defined by \WaspaaTwo{the novel cost function} \eqref{eq:opt_upperBound} and \Waspaa{the unitarity constraint} \eqref{eq:ortho_constraint_upperBound}
\begin{align}
\underset{\W_{\FreqIdx} \forall \FreqIdx}{\text{minimize}} &\ \frac{1}{2}\sum_{\ChannelIdx=1}^{\ChannelIdxMax}\sum_{\FreqIdx=1}^{\FreqIdxMax}\w_{\ChannelIdx,\FreqIdx}\herm\V_{\ChannelIdx,\FreqIdx}\w_{\ChannelIdx,\FreqIdx} \WalterThree{\overset{c}{=} \UpperBound\left(\SetW\vert\SetW^{(l)}\right)} \label{eq:opt_upperBound}\\
\text{subject to}&\ \W_\FreqIdx\W_\FreqIdx\herm = \mathbf{I}_\ChannelIdxMax \ \forall \FreqIdx, \label{eq:ortho_constraint_upperBound}
\end{align}
with equality of \eqref{eq:opt_upperBound} to \eqref{eq:IVA_cost_function} iff $\SetW = \SetW^{(l)}$. Equality up to a constant is denoted by $\overset{c}{=}$. By inspection of \eqref{eq:opt_upperBound}, we see that the optimization w.r.t. the demixing matrices $\W_{\FreqIdx}$ is now expressed by the optimization of demixing filter vectors $\w_{\ChannelIdx,\FreqIdx}$ separately for different frequency bins and channels. However, the channel-wise demixing filters $\w_{\ChannelIdx,\FreqIdx}$ are coupled within one frequency bin due to the constraint \eqref{eq:ortho_constraint_upperBound}. 
To simplify the problem, we divide the optimization of \eqref{eq:opt_upperBound} and \eqref{eq:ortho_constraint_upperBound} into two steps: \Waspaa{a)} Relaxing the constraint \eqref{eq:ortho_constraint_upperBound}\Waspaa{, which allows for solving} \eqref{eq:opt_upperBound} for each demixing filter $\w_{\ChannelIdx,\FreqIdx}$ \Waspaa{without being influenced by the other demixing filters.} \Waspaa{b)} Imposing \eqref{eq:ortho_constraint_upperBound} by projecting the results from \Waspaa{a)} onto the set of unitary matrices, the so-called complex Stiefel manifold.

For \Waspaa{Step} \Waspaa{a)}, we replace the unitarity constraint \eqref{eq:ortho_constraint_upperBound} by a unit norm constraint for the demixing filters and obtain an optimization problem which is now only dependent on a single output channel $\ChannelIdx$
\begin{align}
\underset{\w_{\ChannelIdx,\FreqIdx}}{\text{minimize}} &\quad \frac{1}{2}\w_{\ChannelIdx,\FreqIdx}\herm \V_{\ChannelIdx,\FreqIdx}\w_{\ChannelIdx,\FreqIdx}\label{eq:upper_bound}\\
\text{subject to}&\quad \Vert\w_{\ChannelIdx,\FreqIdx}\Vert_2^2 = 1.
\label{eq:upper_bound_constraint}
\end{align}
Optimization by using the Lagrangian multiplier $\lambda_{\ChannelIdx,\FreqIdx}$ yields the following eigenvalue problem
\begin{equation}
\V_{\ChannelIdx,\FreqIdx}\w_{\ChannelIdx,\FreqIdx} = \lambda_{\ChannelIdx,\FreqIdx}\w_{\ChannelIdx,\FreqIdx},
\label{eq:EV_equation}
\end{equation}
which shows that the eigenvalues of $\V_{\ChannelIdx,\FreqIdx}$ are the critical points of the optimization problem \eqref{eq:upper_bound}, \eqref{eq:upper_bound_constraint}. By multiplication of \eqref{eq:EV_equation} with $\w_{\ChannelIdx,\FreqIdx}\herm$ from the left, we obtain
\begin{equation}
\w_{\ChannelIdx,\FreqIdx}\herm\V_{\ChannelIdx,\FreqIdx}\w_{\ChannelIdx,\FreqIdx} = \lambda_{\ChannelIdx,\FreqIdx}\w_{\ChannelIdx,\FreqIdx}\herm\w_{\ChannelIdx,\FreqIdx} = \lambda_{\ChannelIdx,\FreqIdx}.
\end{equation}
Hence, the optimal $\w_{\ChannelIdx,\FreqIdx}$ is the eigenvector of $\V_{\ChannelIdx,\FreqIdx}$ corresponding to the smallest eigenvalue $\lambda_{\ChannelIdx,\FreqIdx}$ (as $\V_{\ChannelIdx,\FreqIdx}$ is Hermitian, its eigenvalues are real-valued and can be ordered). If the smallest eigenvalue $\lambda_{\ChannelIdx,\FreqIdx}$ has algebraic multiplicity one, the choice of $\w_{\ChannelIdx,\FreqIdx}$ is unique up to an arbitrary phase term, i.e., all elements of the set
\begin{equation}
\left\lbrace\w_{\ChannelIdx,\FreqIdx}\in \mathbb{C}^\ChannelIdxMax\vert \w_{\ChannelIdx,\FreqIdx}=e^{j\phi}\tilde{\w}_{\ChannelIdx,\FreqIdx}\right\rbrace,
\end{equation}
where $\phi$ denotes an arbitrary phase and $\tilde{\w}_{\ChannelIdx,\FreqIdx}$ is a solution of \eqref{eq:upper_bound} and \eqref{eq:upper_bound_constraint}, represent equivalent solutions. Under the natural assumption of distinct temporal variance patterns of the source signals, the eigenvalues of $\V_{\ChannelIdx,\FreqIdx}$ can be assumed to be distinct and, hence, the solution for $\w_{\ChannelIdx,\FreqIdx}$ is unique up to an arbitrary phase term.

For Step \Waspaa{b)}, i.e., to impose the unitarity constraint \eqref{eq:ortho_constraint_upperBound} on the demixing matrices $\tilde{\W}_\FreqIdx$ obtained from collecting the demixing filter vectors from Step \Waspaa{a)}, the closest unitary matrix in terms of the Frobenius \Waspaa{distance} is calculated
\begin{equation}
\W_\FreqIdx = \underset{\mathbf{T}_\FreqIdx\in\mathcal{O}_{\ChannelIdxMax\times\ChannelIdxMax}}{\mathrm{argmin}}\Vert \tilde{\W}_\FreqIdx-\mathbf{T}_\FreqIdx\Vert_\text{F}^2,
\end{equation}
where $\mathcal{O}_{\ChannelIdxMax\times\ChannelIdxMax}$ denotes the set of $\ChannelIdxMax\times\ChannelIdxMax$ unitary matrices. This results in \cite{philippe_algorithm_1987}
\begin{equation}
\W_\FreqIdx = \left(\tilde{\W}_\FreqIdx\tilde{\W}_\FreqIdx\herm\right)^{\text{-}\frac{1}{2}}\tilde{\W}_\FreqIdx.
\label{eq:orthogonalization}
\end{equation}
The \ac{MM} algorithm alternates now between two steps: construction of the upper bound by parameterization of \Waspaa{the proposed surrogate optimization problem \eqref{eq:opt_upperBound}, \eqref{eq:ortho_constraint_upperBound}} with the weighted covariance matrix $\V_{\ChannelIdx,\FreqIdx}$ (see \eqref{eq:weighted_covMat}) and minimization of it by calculating the demixing filters $\w_{\ChannelIdx,\FreqIdx}$ by eigenvalue decomposition and orthogonalization of the demixing matrices \eqref{eq:orthogonalization}. This is summarized in Alg.~\ref{alg:pseudocode}.
\begin{algorithm}
	\caption{Faster\ac{IVA}}
	\label{alg:pseudocode}
	\begin{algorithmic}
		\STATE \textbf{INPUT:} Microphone signals $\x_{\FreqIdx,\BlockIdx}$ $\forall \FreqIdx,\BlockIdx$
		\STATE \textbf{Whitening:} Estimate $\WhiteningMat_\FreqIdx$ $\forall \FreqIdx$ and $\xWhite_{\FreqIdx,\BlockIdx} = \WhiteningMat_\FreqIdx\x_{\FreqIdx,\BlockIdx}$ $\forall \FreqIdx,\BlockIdx$ 
		\STATE \textbf{Initialize:} $\W_\FreqIdx^{(0)} = \mathbf{I}_\ChannelIdxMax$ $\forall \FreqIdx$ and $\y_{\FreqIdx,\BlockIdx}^{(0)} = \xWhite_{\FreqIdx,\BlockIdx}$ $\forall \FreqIdx,\BlockIdx$
		\FOR{$l=1$ \TO $L$}
		\STATE $r_{\ChannelIdx,\BlockIdx} = \Vert \yFreqVec_{\ChannelIdx,\BlockIdx}^{(l)} \Vert_2$ $\forall \ChannelIdx,\BlockIdx$
		\FOR{$\FreqIdx=1$ \TO $\FreqIdxMax$}
		\FOR{$\ChannelIdx=1$ \TO $\ChannelIdxMax$} 
		\STATE Estimate $\V_{\ChannelIdx,\FreqIdx}$ by \eqref{eq:weighted_covMat}
		\STATE Compute eigenvector $\w_{\ChannelIdx,\FreqIdx}^{(l)}$ corresponding to smallest eigenvalue $\lambda_{\ChannelIdx,\FreqIdx}^{(l)}$ of $\V_{\ChannelIdx,\FreqIdx}$
		\ENDFOR
		\STATE $\W_\FreqIdx^{(l)} \leftarrow \left(\W_\FreqIdx^{(l)}(\W_\FreqIdx^{(l)})\herm\right)^{\text{-}\frac{1}{2}}\W_\FreqIdx^{(l)}$ (see \eqref{eq:orthogonalization})
		\STATE $\y_{\FreqIdx,\BlockIdx}^{(l)} = \W_\FreqIdx^{(l)} \xWhite_{\FreqIdx,\BlockIdx}$
		\ENDFOR
		\ENDFOR
		\STATE Backprojection
		\STATE \textbf{OUTPUT:} Demixed signals $\y_{\FreqIdx,\BlockIdx}^{(L)}$ $\forall \FreqIdx,\BlockIdx$
	\end{algorithmic}
\end{algorithm}

\section{Experiments}
\label{sec:experiments}
\begin{figure*}[h]
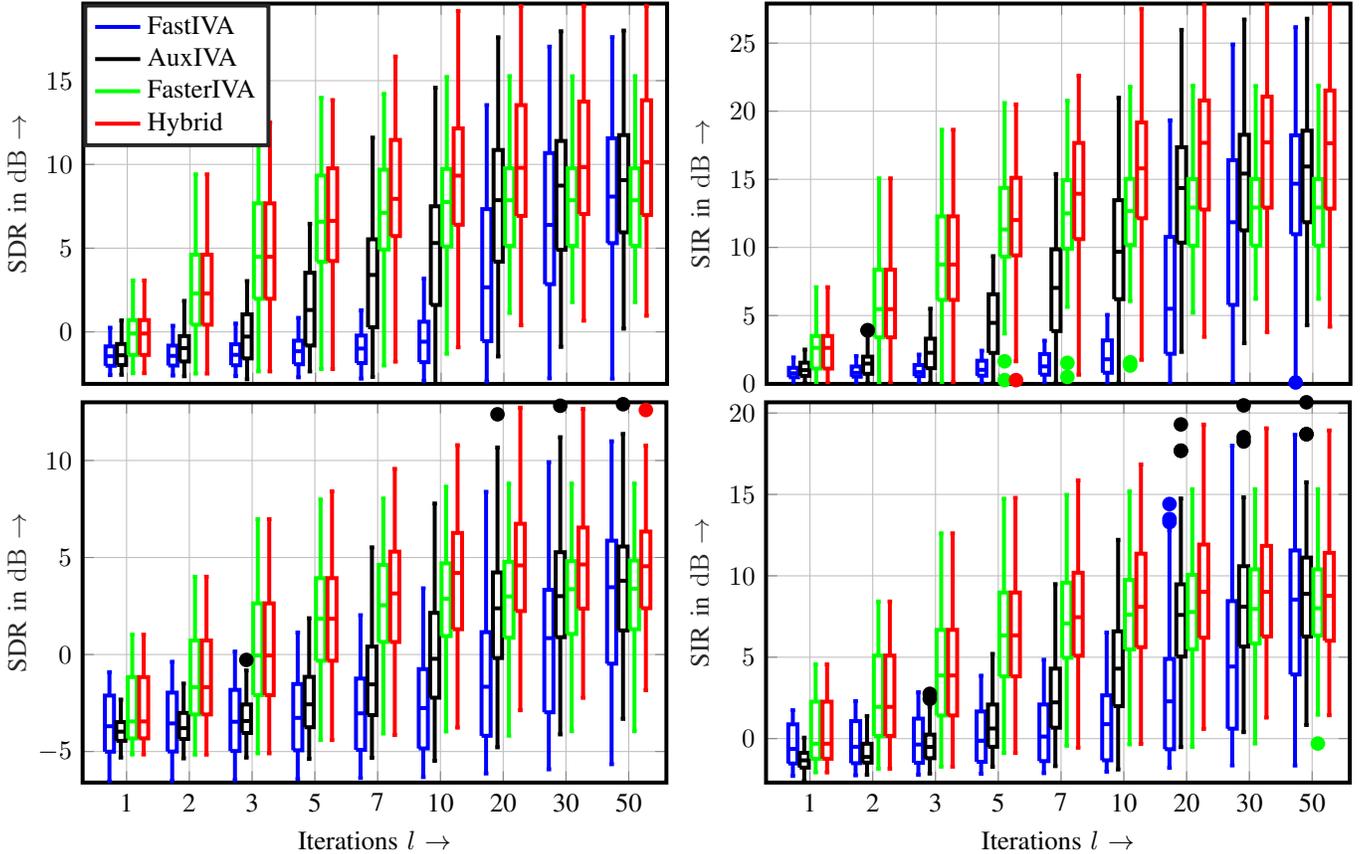

	\setlength\fwidth{0.45\textwidth}
	\input{figures/fast_box_SDR2}\hspace{5pt}\input{figures/fast_box_SIR2}\vspace{-3pt}\\\input{figures/fast_box_SDR3}\hspace{5pt}\input{figures/fast_box_SIR3}\vspace{-3pt}
	\caption{Performance of the investigated algorithms in terms of \ac{SDR} and \ac{SIR} improvement over the number of iterations. To create the plots, results for three different rooms ($T_{60} = 0.2\,\mathrm{sec},0.4\,\mathrm{sec},0.9\,\mathrm{sec}$) and two different source-array distances ($1\,\mathrm{m}$, $2\,\mathrm{m}$) have been used. Each of these experiments have been repeated 20 times, where the source signals have been chosen from a set of four male and four female speakers. The first row shows results for a determined scenario with two speakers, the second row a scenario with three speakers.}
	\label{fig:results}
\end{figure*}

For the experimental evaluation \Waspaa{of the new negentropy-based \ac{IVA} algorithm relative to other \ac{IVA} algorithms,} we convolved speech signals of $10\,\mathrm{sec}$ length with \acp{RIR} measured in three different rooms: two meeting rooms ($T_{60} \in \{0.2\,\mathrm{s},0.4\,\mathrm{s}\}$) and a seminar room ($T_{60} = 0.9\,\mathrm{s}$). For the measurements, we used a
linear microphone array with $4.2\,\mathrm{cm}$ spacing. To obtain \Walter{representative} results, we considered different measurement setups: source positions at $1\,\mathrm{m}$ and $2\,\mathrm{m}$ distance from the microphone array and at $40^\circ/140^0$ and $40^\circ/90^\circ/140^\circ$ w.r.t. the microphone array axis. To simulate microphone noise, we added white Gaussian noise to the observed signals to obtain a \ac{SNR} of $30\,\mathrm{dB}$. To address the effect of \Walter{source variability}, we chose the clean source signals randomly from a set of four male and four female speakers \Waspaa{and repeated the experiments $20$ times in this way}. The simulated microphone signals are transformed into the \ac{STFT} domain by a Hamming window of length $2048$ and $50\%$ overlap at a sampling frequency of $16\,\mathrm{kHz}$. The performance of the investigated algorithms is measured in terms of \acf{SDR}, \acf{SIR} and \acf{SAR} \cite{vincent_performance_2006}. These performance measures are not directly connected to the cost function, but are closely related to the separation performance as experienced by a human listener. We used for all algorithms a Laplacian source model yielding \mbox{$\tilde{G}(r_{\ChannelIdx,\BlockIdx}) = r_{\ChannelIdx,\BlockIdx}$} \mbox{\cite{kim_blind_2007,ono_stable_2011}}. The scaling ambiguity of the frequency bin-wise estimates is resolved by the backprojection technique \cite{ono_stable_2011}.

To benchmark the results, we compared the performance with two state-of-the-art algorithms: AuxIVA \cite{ono_stable_2011}, which can be considered as the \Waspaa{best performing algorithm} in the field\Waspaa{,} and \mbox{FastIVA} \cite{lee_fast_2007}, which is based on the same cost function as the proposed method \eqref{eq:IVA_cost_function}, but uses a fixed-point algorithm for optimization. Results for the comparison of the investigated algorithms with \ac{IVA} optimized by a natural gradient update scheme \cite{kim_blind_2007} \Waspaa{are} not shown here as its convergence turned out to be exceedingly \Waspaa{slow} and the final values are not better than for the competing methods. \WaspaaTwo{Note that a variation of experimental parameters such as \ac{STFT} length, noise type, \ac{SNR} etc. affected the discussed algorithms similarly.}
The restriction to unitary demixing matrices is well known to yield a \Waspaa{fast initial} convergence at the cost of \Waspaa{inferior steady-state} performance \Waspaa{relative to} methods that require only invertible demixing matrices \cite{cardoso_performance_1994}. Hence, a natural idea is to use the proposed method 'FasterIVA' until \Waspaa{reaching the steady state} and then relax the unitarity constraint by switching to the AuxIVA update rules \WaspaaTwo{(found to be superior to FastIVA in preliminary experiments)} which do not constrain the demixing matrices to be unitary. The switching of this hybrid \Waspaa{approach} from FasterIVA to AuxIVA is triggered once FasterIVA reached a steady state characterized by only small changes of $\W_\FreqIdx$:
\begin{equation}
\frac{1}{\FreqIdxMax \ChannelIdxMax^2}\sum_{\FreqIdx=1}^{\FreqIdxMax}\left\Vert \W_\FreqIdx^{(l-1)} - \W_\FreqIdx^{(l)}\right\Vert_{\text{F}}^2 < \WalterTwo{\gamma}.
\end{equation}
\Waspaa{The} threshold $\gamma$ is chosen here to $\gamma = 0.05$.
The experimental results including all three different rooms, both source-array distances and 20 repeated draws of source signals resulting in $120$ different experimental conditions for each number of sources are shown in Fig.~\ref{fig:results} over the number of iterations. The slowest convergence \Waspaa{among} the discussed methods is obtained by FastIVA. \Waspaa{Often, this} algorithm did \Waspaa{not even reach the steady state} \WalterTwo{within the given number of} iterations. However, \WalterTwo{even after convergence} its \WalterTwo{final values} were still \WalterTwo{not better} than the competing methods in the vast majority of cases. The \ac{MM}-based AuxIVA algorithm outperformed FastIVA in terms of convergence speed but also w.r.t. its final values. The proposed method FasterIVA showed much faster \WalterTwo{initial} convergence \WalterTwo{than both FastIVA and AuxIVA} and usually reached its steady state already after about five iterations. On the other hand, its final values \WalterTwo{were} slightly worse than AuxIVA for the two-source scenarios, while it was the same for the three-source case. The 'Hybrid' approach, which switches after convergence of FasterIVA to the AuxIVA update rules, obtained in all scenarios the fastest convergence and the best final values at a comparable runtime. The values for \ac{SAR} improvement have been omitted here due to a lack of space, but they showed comparable results for the discussed methods with a slight advantage for the hybrid approach.
The runtime per iteration of the investigated methods, which is comparable in most cases, is given in Tab.~\ref{tab:runtime}. \WalterTwo{Note that, e.g., in the 3-source experiment, FasterIVA needs only 4 iterations to reach the $\Delta$\ac{SIR} value of FastIVA after 30 iterations, so that the complexity gain for FasterIVA for comparable performance amounts to a factor of approximately $5$.}
\begin{table}
	\centering
	\begin{tabular}{c|cccc}
		&FastIVA  &AuxIVA  &FasterIVA  &Hybrid  \\ 
		\toprule 
		2 Sources&$1$  &$1$  &$1.36$  &$1.07$  \\
		3 Sources&$1$  &$1.53$  &$1.76$  &$1.59$  \\ 
		\bottomrule 
	\end{tabular}
	\caption{\WalterThree{Runtime per iteration} \Walter{relative to FastIVA}.}
	\label{tab:runtime}
\end{table}
\section{Conclusion}
\label{sec:conclusion}
In this contribution, we presented a fast converging update scheme based on the \ac{MM} principle and an \ac{EVD} of weighted microphone \WalterTwo{sample} covariance matrices. The proposed update scheme outperformed state-of-the-art optimization methods in terms of convergence speed as well as \WalterTwo{final \Waspaa{steady-state} values}.
\WalterThree{As a promising next step}, these update rules \WalterTwo{could} be investigated w.r.t. their efficacy for the optimization of \ac{ILRMA}-type algorithms.
\bibliographystyle{IEEEbib}
\bibliography{refs}

\end{document}